\documentclass[letter]{aa} 
\usepackage{graphicx}
\usepackage{txfonts}
\usepackage{natbib}
\def\hho{H$_2$O}

\def\hh{H$_2$}

\def\cp{C$^+$}
\def\mic{$\mu$m}
\def\kms{km\,s$^{-1}$}
\def\kkms{K\,km\,s$^{-1}$}
\def\pow#1#2{#1$\times$10$^{#2}$}
\def\rcm{cm$^{-1}$}
\def\scm{cm$^{-2}$}
\def\ccm{cm$^{-3}$}

\def\dv{$\Delta${\it V}}
\def\vlsr{$V_{\rm LSR}$}

\def\ltsim{{_<\atop{^\sim}}}

\def\tmb{$T_{\rm mb}$}
\def\new#1{{#1}}
\def\newnew#1{{#1}}
\begin{document}
\title{Detection of HF emission from the Orion Bar
\thanks{\textit{Herschel} is an ESA space observatory with science instruments provided
by European-led Principal Investigator consortia and with important participation from NASA}}
\titlerunning{HF emission from the Orion Bar}

\author{F.F.S. van der Tak \inst{\ref{sron},\ref{rug}} \and
        V. Ossenkopf \inst{\ref{cologne}} \and
        Z. Nagy \inst{\ref{sron},\ref{rug}} \and
        A. Faure \inst{\ref{grenoble}} \and
        M. R\"ollig \inst{\ref{cologne}} \and
        E.A. Bergin \inst{\ref{michigan}} 
}

\institute{SRON Netherlands Institute for Space Research, Landleven 12, 9747 AD Groningen, The Netherlands; \email{vdtak@sron.nl} \label{sron} \and
  Kapteyn Institute, University of Groningen, The Netherlands \label{rug} \and
  KOSMA, I. Physik. Inst., Universit\"at zu K\"oln, Germany \label{cologne} \and
  UJF-Grenoble~1/CNRS, Institut de Plan\'etologie et d'Astrophysique de Grenoble, France \label{grenoble} \and
  Dept. of Astronomy, University of Michigan, Ann Arbor, USA \label{michigan}
          }

\date{Submitted 24 November 2011; accepted 15 December 2011}

\abstract
{The clumpy density structure of photon-dominated regions is well established, but the physical properties of the clumps and of the surrounding interclump medium are only approximately known.}
{The aim of this paper is to constrain the physical and chemical conditions in the Orion Bar, a prototypical nearby photon-dominated region.}
{We present observations of the HF $J$=1--0 line, which appears in emission toward the Orion Bar, and compare the brightness of the line to non-LTE radiative transfer calculations. 
}
{The large width of the HF line suggests an origin of the emission in the interclump gas, but collisional excitation by \hh\ in the interclump gas underpredicts the observed line intensity by factors of 3--5. 
In contrast, an origin of the line in the dense clumps requires a density of $\sim$10$^9$\ccm, 10--100 times higher than previous estimates, which is unlikely. 
However, electron impact excitation reproduces our observations for $T$=100\,K and $n_e$=10\,\ccm, as expected for the interclump gas.}
{We conclude that HF emission is a signpost of molecular gas with a high electron density. Similar conditions may apply to active galactic nuclei where HF also appears in emission.}


\maketitle

\section{Introduction}
\label{s:intro}


Photon-dominated regions (PDRs) are the surface regions of molecular clouds, where ultraviolet radiation with photon energies between a few and 13.6\,eV drives the thermal and chemical balance of the gas \citep{hollenbach1999}.
This situation occurs in regions of high-mass star formation, but also in protoplanetary disks and in the nuclei of active galaxies. Studying the structure of PDRs therefore has a wide astrophysical application.
In PDRs, gas heating proceeds by photo-electric emission from dust grains, while the main cooling channels are the fine structure lines of C$^+$ and O and the rotational lines of CO \citep{kaufman1999}.
Absorption of the impinging ultraviolet radiation by dust and gas in the PDR creates a layered structure, where chemical transitions such as H$^+$ $\to$ H $\to$ \hh\ and C$^+$ $\to$ C $\to$ CO occur.

The Orion Bar is a prototypical PDR, located between the Orion molecular cloud and the Orion Nebula, the H$^+$ region surrounding the Trapezium stars, at a distance of 414\,pc \citep{menten2007}. Observations at infrared and submillimeter wavelengths indicate a geometry for the Bar where the PDR is wrapped around the Orion Nebula, and changes from a face-on to an edge-on view where the molecular emission peaks (\citealt{hogerheijde1995}; \citealt{walmsley2000}).
The mean temperature of the molecular gas in the Bar is 85\,K, while the temperature rises to several 100\,K toward the ionization front, where the emission from PAH particles and vibrationally excited \hh\ peaks.

While the temperature structure of the Orion Bar is reasonably well understood, the same cannot be said about the density structure. The mean density of the molecular gas is 10$^5$\,\ccm, but single-dish observations already indicate the presence of small-scale density variations without apparent pattern, usually called ''clumps'' \citep{hogerheijde1995}, that are also seen toward other PDRs \citep{stutzki1988,wang1993}. While interferometric observations have confirmed the presence of clumps \citep{youngowl2000}, the densities of both the clumps and the interclump medium are somewhat uncertain. The interclump medium probably has a density between a few 10$^4$ and \pow{2}{5}\,\ccm\ \citep{simon1997}, while estimates of the clump density range from \pow{1.5}{6} to \pow{6}{6}\,\ccm\ \citep{lis2003}.

This Letter presents observations of the HF $J$=1--0 line, which appears in emission toward the Orion Bar.
The HF molecule is expected to be the dominant carrier of gas-phase fluorine, because the reaction F + \hh\ $\to$ HF + H is exothermic. 
For diffuse clouds where the effect of depletion on grains should be unimportant, models by \citet{neufeld2009} predict an HF abundance of $\sim$\pow{3.6}{-8} relative to \hh.
Recent observations of the HF 1--0 line confirm this prediction: the line is seen in absorption toward several background sources, indicating abundances of $\sim$\pow{1--2}{-8} \citep{neufeld2010}.
Toward dense clouds, the abundance is measured to be $\sim$100 times lower \citep{phillips2010}, suggesting that depletion of F on grain surfaces or excitation effects play a role.

Like CO, HF is a linear rotor with a regular line spectrum, where $\nu = 2B(J+1)$ and $A_{ij} \propto \nu^3$.
Unlike CO, HF has a small reduced mass and a large dipole moment, so that the lines have high frequencies and radiative decay is rapid.
In particular, the HF 1--0 line has a frequency of 1232\,GHz and an Einstein~A coefficient of \pow{2.422}{-2}\,s$^{-1}$. 
Thermal excitation of the $J$=1 level of HF thus requires extremely high gas densities, which is why the line usually appears in absorption \citep{sonnentrucker2010,monje2011}.

\section{Observations and data reduction}
\label{s:obs}

\begin{table}
\caption{Line parameters.}
\label{t:pars}
\begin{tabular}{cccccc}
\hline \hline
\noalign{\smallskip}
$\int$ \tmb\ \dv\ & \vlsr\ & \dv\  & \tmb\  \\
\kkms\                 &  \kms\  & \kms\  & K  \\ 
\noalign{\smallskip}
\hline
\noalign{\smallskip}
7.79(21)             & 9.23(7) & 4.86(15) & 1.50 \\
\noalign{\smallskip}
\hline
\noalign{\smallskip}
\end{tabular}
\tablefoot{Numbers in parentheses are error bars in units of the last decimal.}
\end{table}

The observations were made with the Heterodyne Instrument for the Far-Infrared (HIFI; \citealt{degraauw2010}) onboard ESA's \textit{Herschel} Space Observatory \citep{pilbratt2010} on 2011 March 19, as part of the Guaranteed Time key program \textit{Herschel Observations of Extraordinary Sources} (HEXOS; \citealt{bergin:hexos}).
Receiver band 5b was used as front end and the acousto-optical Wide-Band Spectrometer (WBS), 
which covers 4\,GHz bandwidth in four 1140\,MHz subbands at 1.1\,MHz (0.30\,\kms) resolution, as back end.
The data were taken in spectral survey mode with a redundancy of 6, using the frequency switch technique with a throw of 100\,MHz.

The position observed is the CO$^+$ peak of the Orion Bar at $\alpha$ = 05$^h$ 35$^m$ 20.6$^s$, $\delta$ = --05$^\circ$ 25$'$ 14$''$ (J2000) \citep{stoerzer1995}.
The FWHM beam size at the observing frequency is 17.3 arcsec \citep{roelfsema2011}, which corresponds to 7200\,AU at the distance of Orion.
The system temperature for these data is 2571\,K in SSB main beam units, and the integration time is 70 minutes (ON+OFF).

Calibration of the data, removal of standing waves, and sideband deconvolution were performed in the \textit{Herschel} Interactive Processing Environment (HIPE) version 6.0.
The subsequent analysis was performed within the CLASS\footnote{\tt http://www.iram.fr/IRAMFR/GILDAS} package, version January 2011. 
The calibration is estimated to be accurate to $\approx$10\%; the velocity scale should be accurate to 0.5\,\kms\ or better.
The intensity scale was converted to \tmb\ using a main beam efficiency of 64\% \citep{roelfsema2011}. 
Only a linear baseline was subtracted.
After inspection, data from the two polarizations were averaged together to obtain an rms noise level of 0.18\,K in \tmb\ units on 0.5\,MHz channels. 

\section{Results}
\label{s:res}

\begin{figure}[tb]
\centering
\includegraphics[width=7cm,angle=-90]{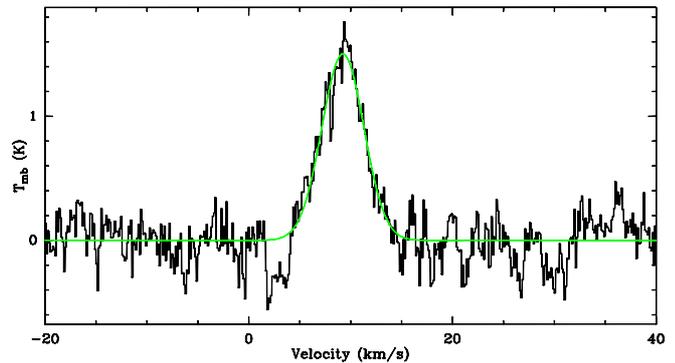}
\caption{Spectrum of the HF $J$=1--0 line, observed with Herschel-HIFI toward the Orion Bar.}
\label{f:profi}
\end{figure}

Figure~\ref{f:profi} presents the HF line profile observed toward the Orion Bar. Strong emission is detected at \vlsr\ = +9.24\,\kms, with an almost Gaussian shape. In addition, weak absorption may be present near \vlsr\ = +2..+4 \kms, which appears in both polarizations. If confirmed, this absorption arise in the Orion Ridge \citep{tauber1994}.
Standing waves prohibit the detection of  continuum signal in our data down to an rms of $\approx$1\,K in $T_A^*$.

Table~\ref{t:pars} lists the parameters of the emission feature, measured by fitting a Gaussian profile to the data in Figure~\ref{f:profi}. The measured \vlsr\ agrees well with values for the dense molecular gas in the Orion Bar as probed by H$_2$CO and CH$_3$OH lines in the 100--400\,GHz range (\citealt{hogerheijde1995}; \citealt{leurini2006}). However, the FWHM line width of 4.86\,\kms\ is much broader than measured for the dense molecular gas, by a factor of $\ga$2. This difference is about twice as large as expected from the lower mass of the HF molecule compared with species such as CO. 
The width of the HF line is close to the values measured for C$^+$ and CH$^+$ (Nagy et al, in prep.), suggesting an origin of the observed HF emission in the low-density interclump gas of the Orion Bar, which is a surprise given the large critical density ($\sim$\pow{5}{9}\,\ccm) of the line.

\section{Excitation by \hh\ }
\label{s:therm}

To clarify the origin of the observed HF emission, we have performed model calculations with the RADEX program \citep{vdtak2007}.
This program solves the equations of statistical equilibrium in the presence of collisional and radiative excitation, treating optical depth effects in the lines with an escape probability formalism. 
%
%
The calculations use spectroscopic input data from the JPL catalog \citep{pickett1998}, and collisional rate coefficients for the HF-\hh\ system from \citet{guillon2011}.
These collision rates are $\approx$10 times larger than calculated for the HF-He system \citep{reese2005}, even after scaling for the difference in reduced mass \citep{schoeier2005}. This difference is at the upper end of the expected range \citep{vdtak2011}, due to the large well depth of the HF-\hh\ interaction potential.
All calculations assume a line width of 5\,\kms\ (Table~\ref{t:pars}) and a background radiation field of 2.73\,K due to the cosmic microwave background.
We assume that the ortho-para ratio of \hh\ equals its thermal value.
The free parameters in the calculation are the kinetic temperature, the gas density $n$(\hh) and the HF column density $N$(HF).

Based on the measured line width, we first consider an origin of the HF emission in the warm interclump gas close to the ionization front. We adopt $n$(\hh) = 10$^5$\,\ccm, $N$(\hh) = 10$^{22}$\,\scm\ \newnew{\citep{hogerheijde1995}}, and assume that all fluorine (at Solar abundance) is in the form of HF, as is commonly observed for diffuse interstellar clouds. This assumption implies an HF abundance of \pow{3.6}{-8}, so that $N$(HF) = \pow{3.6}{14}\,\scm. For a gas kinetic temperature of 100\,K, this model predicts an intensity for the HF 1--0 line of 0.5\,K, which is a factor of 3 lower than observed. 
\new{Models in which the HF emission arises in warm surface layers with a low molecular fraction and accordingly a low HF abundance also do not match our observations, provided that HF-H collision rates are similar to those for HF-\hh.}
We conclude that the observed HF emission does not originate in the warm interclump gas if collisional excitation by \hh\ \new{or H} is responsible.

As an alternative model, we consider an origin of the HF emission in the dense clumps of the Orion Bar. For this gas, we assume a kinetic temperature of 85\,K \citep{hogerheijde1995}, $N$(\hh) $\approx$\pow{1}{23} (e.g., \citealt{vdwiel2009}) and an HF abundance of $\ltsim$\pow{3.6}{-8}. 
These models reproduce the observed HF line intensity for gas densities in the range 10$^8$--10$^9$\,\ccm\ \newnew{and $N$(HF) in the range 10$^{12}$--10$^{13}$\,\scm. }
The line optical depth in these models is around unity, which is consistent with the observed line shape. 
The implied HF abundance is $\sim$10$^{-10}$--10$^{-11}$, which is similar to the values found for dense gas in previous studies \citep{phillips2010}.
\newnew{Since the line is effectively optically thin, lower \hh\ volume densities require correspondingly higher HF column densities, so that the elemental fluorine abundance would be exceeded.}
We conclude that thermal excitation of the HF line indicates a high gas density of $n$=10$^8$--10$^9$\,\ccm, similar to the inner envelopes of late-type stars where HF emission is also seen \citep{agundez2011}.

One problem with this alternative model is that the required gas density of $\sim$10$^9$\,\ccm\ is much higher than commonly adopted for the Orion Bar. An origin of the HF emission in high-density clumps is also inconsistent with the line width, which is much larger than that of CO. In addition, while peak densities of $\sim$10$^9$\,\ccm\ may occur locally, this value is unreasonably high as an average over the 7200\,AU HIFI beam. Finally, the Orion Bar does not stand out for its high density between regions where HF has been observed, which all show the line in absorption.


\section{Excitation by electrons }
\label{s:elec}

Although collisional excitation of HF by \hh\ does not match our observations of the Orion Bar, \hh\ is not the only possible collision partner responsible for the excitation of HF in PDR-type regions. In the interclump gas of the Orion Bar, most carbon is in the form of C$^+$, leading to an electron abundance of $\sim$10$^{-4}$ and an electron density of 10\,\ccm. Under these conditions, a model without collisional excitation by electrons would be incomplete. Quantum mechanical calculations of collisional cross sections for the e-HF system only exist for $T > 500$K \citep{thuemmel1992}, but for highly polar molecules such as HF, the Born approximation (see, e.g., \citealt{itikawa2005}) should give reliable results as long as $kT > E_{\rm up}$. 
\new{We have calculated electron impact excitation rates for the $\Delta${\it J} = 1 transitions of HF in the Born approximation, and scaled the values to the results of the R-matrix calculations by \citet{thuemmel1992}. These rates should be accurate to within a factor of $\approx$2, similar to the well-studied e-\hho\ system with a similar dipole moment \citep{zhang2009}, and sufficient for our purposes. Calculations with RADEX indicate that for $T = 100$\,K and $n_e = 10$\,\ccm, the observed HF line intensity is reproduced for $N$(HF)$\approx$\pow{1}{15}\,\scm.}

\new{The corresponding \hh\ column density may be estimated from the \cp\ observations by \citet{herrmann1997} with the KAO in a 55$''$ beam. These data indicate that $N$(\cp) $\approx$\pow{(3--5)}{18}\,\scm, so that for \cp/\hh\ = 2$\times$C/H = \pow{2.4}{-4} \citep{wakelam2008,sofia2011}, we obtain $N$(\hh) = \pow{(1.3--2.0)}{22}\,\scm. This value is a lower limit because of beam dilution: mapping of \cp\ with HIFI indicates that $N$(\cp) and therefore $N$(\hh) is $\sim$3--5 times higher (R{\"ollig et al, in prep.}). The HF abundance in the \cp--\hh\ layer is thus \pow{(3--4)}{-8}, consistent with the Solar fluorine abundance.}
We conclude that electron excitation in the interclump medium is a viable model to explain the observed HF emission in the Orion Bar.

\section{Conclusions}
\label{s:concl}

We have presented observations of the HF $J$=1--0 line toward the Orion Bar, which is the first time that this line is seen in emission from the Galactic interstellar medium. We present calculations and arguments why this emission cannot be caused by collisional excitation of HF by \hh, either from dense clumps or from the inter-clump medium. 
However, collisional excitation by electrons in the interclump gas explains our observations. The appearance of HF in emission therefore seems to be a signpost of molecular gas with a high electron density, which in the case of the Orion Bar is the combined effect of the high gas density and the strong ultraviolet radiation field. 

The physical conditions of the Orion Bar may be similar to those in the nucleus of the active galaxy Mrk 231, where the radiation field is strong and where the HF line also appears in emission \citep{vdwerf2010}. 
Other active galactic nuclei where HF appears in absorption such as Arp 220 \citep{rangwala2011} and the Cloverleaf \citep{monje2011clover} probably have lower electron densities, which may indicate a softer radiation field. 

Support for an origin of the HF emission in the interclump medium comes from the detection of CF$^+$ toward the Orion Bar \citep{neufeld2006}. Additional support for this model would come from observations of H$_2$F$^+$, for which line frequencies are known \citep{fujimori2011}. However, our data do not show any features at the predicted line frequencies down to an rms of $\approx$50\,mK in $T_A^*$, which is a factor of $\sim$10 above the expected signal.

In the future, maps of HF emission with Herschel-HIFI will help to clarify the spatial distribution of dense gas clumps in the Orion Bar and other PDRs.
Another useful test would be the observation of HF $J$=2--1 or higher-$J$ lines in the $v$=0 or $v$=1 states. Unfortunately, the $J$=2--1 line at 2463\,GHz lies very close (0.13\,\mic) to the N$^+$ $^3P_2$--$^3P_1$ fine-structure line at 121.8\,\mic, so that the lines are blended at the resolution of the PACS instrument onboard Herschel.
Heterodyne observations in this frequency range may become possible with the future STO telescope, the SOFIA airborne observatory, or the Millimetron mission.

\begin{acknowledgements}

The authors thank Frank Helmich for useful discussions, 
Gr\'egoire Guillon and Thierry Stoecklin for sending their HF-\hh\ collision data ahead of publication,
Yunhee Choi for help with the data reduction,
and John Black and David Neufeld for comments on the manuscript.
Part of this work was supported by the \emph{Deut\-sche
For\-schungs\-ge\-mein\-schaft} through grant SFB 956C.

HIFI has been designed and built by a consortium of institutes and university departments from across Europe, Canada and the US under the leadership of SRON Netherlands Institute for Space Research, Groningen, The Netherlands with major contributions from Germany, France and the US. Consortium members are: Canada: CSA, U.Waterloo; France: CESR, LAB, LERMA, IRAM; Germany: KOSMA, MPIfR, MPS; Ireland, NUI Maynooth; Italy: ASI, IFSI-INAF, Arcetri-INAF; Netherlands: SRON, TUD; Poland: CAMK, CBK; Spain: Observatorio Astron\'omico Nacional (IGN), Centro de Astrobiolog\'{\i}a (CSIC-INTA); Sweden: Chalmers University of Technology - MC2, RSS \& GARD, Onsala Space Observatory, Swedish National Space Board, Stockholm University - Stockholm Observatory; Switzerland: ETH Z\"urich, FHNW; USA: Caltech, JPL, NHSC.
\end{acknowledgements}

\bibliographystyle{aa}
\bibliography{hf}

\begin{thebibliography}{41}
\expandafter\ifx\csname natexlab\endcsname\relax\def\natexlab#1{#1}\fi

\bibitem[{{Ag{\'u}ndez} {et~al.}(2011){Ag{\'u}ndez}, {Cernicharo}, {Waters},
  {Decin}, {Encrenaz}, {Neufeld}, {Teyssier}, \& {Daniel}}]{agundez2011}
{Ag{\'u}ndez}, M., {Cernicharo}, J., {Waters}, L.~B.~F.~M., {et~al.} 2011,
  \aap, 533, L6

\bibitem[{{Bergin} {et~al.}(2010){Bergin}, {Phillips}, {Comito}, {Crockett},
  {Lis}, {Schilke}, {Wang}, {Bell}, {Blake}, {Bumble}, {Caux}, {Cabrit},
  {Ceccarelli}, {Cernicharo}, {Daniel}, {de Graauw}, {Dubernet},
  {Emprechtinger}, {Encrenaz}, {Falgarone}, {Gerin}, {Giesen}, {Goicoechea},
  {Goldsmith}, {Gupta}, {Hartogh}, {Helmich}, {Herbst}, {Joblin}, {Johnstone},
  {Kawamura}, {Langer}, {Latter}, {Lord}, {Maret}, {Martin}, {Melnick},
  {Menten}, {Morris}, {M{\"u}ller}, {Murphy}, {Neufeld}, {Ossenkopf}, {Pagani},
  {Pearson}, {P{\'e}rault}, {Plume}, {Roelfsema}, {Qin}, {Salez}, {Schlemmer},
  {Stutzki}, {Tielens}, {Trappe}, {van der Tak}, {Vastel}, {Yorke}, {Yu}, \&
  {Zmuidzinas}}]{bergin:hexos}
{Bergin}, E.~A., {Phillips}, T.~G., {Comito}, C., {et~al.} 2010, \aap, 521, L20

\bibitem[{{De Graauw} {et~al.}(2010){De Graauw}, {Helmich}, {Phillips},
  {Stutzki}, {Caux}, {Whyborn}, {Dieleman}, {Roelfsema}, {Aarts}, {Assendorp},
  {Bachiller}, {Baechtold}, {Barcia}, {Beintema}, {Belitsky}, {Benz}, {Bieber},
  {Boogert}, {Borys}, {Bumble}, {Ca{\"i}s}, {Caris}, {Cerulli-Irelli},
  {Chattopadhyay}, {Cherednichenko}, {Ciechanowicz}, {Coeur-Joly}, {Comito},
  {Cros}, {de Jonge}, {de Lange}, {Delforges}, {Delorme}, {den Boggende},
  {Desbat}, {Diez-Gonz{\'a}lez}, {di Giorgio}, {Dubbeldam}, {Edwards},
  {Eggens}, {Erickson}, {Evers}, {Fich}, {Finn}, {Franke}, {Gaier}, {Gal},
  {Gao}, {Gallego}, {Gauffre}, {Gill}, {Glenz}, {Golstein}, {Goulooze},
  {Gunsing}, {G{\"u}sten}, {Hartogh}, {Hatch}, {Higgins}, {Honingh}, {Huisman},
  {Jackson}, {Jacobs}, {Jacobs}, {Jarchow}, {Javadi}, {Jellema}, {Justen},
  {Karpov}, {Kasemann}, {Kawamura}, {Keizer}, {Kester}, {Klapwijk}, {Klein},
  {Kollberg}, {Kooi}, {Kooiman}, {Kopf}, {Krause}, {Krieg}, {Kramer},
  {Kruizenga}, {Kuhn}, {Laauwen}, {Lai}, {Larsson}, {Leduc}, {Leinz}, {Lin},
  {Liseau}, {Liu}, {Loose}, {L{\'o}pez-Fernandez}, {Lord}, {Luinge}, {Marston},
  {Mart{\'{\i}}n-Pintado}, {Maestrini}, {Maiwald}, {McCoey}, {Mehdi}, {Megej},
  {Melchior}, {Meinsma}, {Merkel}, {Michalska}, {Monstein}, {Moratschke},
  {Morris}, {Muller}, {Murphy}, {Naber}, {Natale}, {Nowosielski}, {Nuzzolo},
  {Olberg}, {Olbrich}, {Orfei}, {Orleanski}, {Ossenkopf}, {Peacock}, {Pearson},
  {Peron}, {Phillip-May}, {Piazzo}, {Planesas}, {Rataj}, {Ravera}, {Risacher},
  {Salez}, {Samoska}, {Saraceno}, {Schieder}, {Schlecht}, {Schl{\"o}der},
  {Schm{\"u}lling}, {Schultz}, {Schuster}, {Siebertz}, {Smit}, {Szczerba},
  {Shipman}, {Steinmetz}, {Stern}, {Stokroos}, {Teipen}, {Teyssier}, {Tils},
  {Trappe}, {van Baaren}, {van Leeuwen}, {van de Stadt}, {Visser}, {Wildeman},
  {Wafelbakker}, {Ward}, {Wesselius}, {Wild}, {Wulff}, {Wunsch}, {Tielens},
  {Zaal}, {Zirath}, {Zmuidzinas}, \& {Zwart}}]{degraauw2010}
{De Graauw}, T., {Helmich}, F.~P., {Phillips}, T.~G., {et~al.} 2010, \aap, 518,
  L6

\bibitem[{{Fujimori} {et~al.}(2011){Fujimori}, {Kawaguchi}, \&
  {Amano}}]{fujimori2011}
{Fujimori}, R., {Kawaguchi}, K., \& {Amano}, T. 2011, \apjl, 729, L2

\bibitem[{Guillon \& Stoecklin(2011)}]{guillon2011}
Guillon, G. \& Stoecklin, T. 2011, \mnras, in press

\bibitem[{{Herrmann} {et~al.}(1997){Herrmann}, {Madden}, {Nikola}, {Poglitsch},
  {Timmermann}, {Geis}, {Townes}, \& {Stacey}}]{herrmann1997}
{Herrmann}, F., {Madden}, S.~C., {Nikola}, T., {et~al.} 1997, \apj, 481, 343

\bibitem[{{Hogerheijde} {et~al.}(1995){Hogerheijde}, {Jansen}, \& {van
  Dishoeck}}]{hogerheijde1995}
{Hogerheijde}, M.~R., {Jansen}, D.~J., \& {van Dishoeck}, E.~F. 1995, \aap,
  294, 792

\bibitem[{{Hollenbach} \& {Tielens}(1999)}]{hollenbach1999}
{Hollenbach}, D.~J. \& {Tielens}, A.~G.~G.~M. 1999, Reviews of Modern Physics,
  71, 173

\bibitem[{{Itikawa} \& {Mason}(2005)}]{itikawa2005}
{Itikawa}, Y. \& {Mason}, N. 2005, \physrep, 414, 1

\bibitem[{{Kaufman} {et~al.}(1999){Kaufman}, {Wolfire}, {Hollenbach}, \&
  {Luhman}}]{kaufman1999}
{Kaufman}, M.~J., {Wolfire}, M.~G., {Hollenbach}, D.~J., \& {Luhman}, M.~L.
  1999, \apj, 527, 795

\bibitem[{{Leurini} {et~al.}(2006){Leurini}, {Rolffs}, {Thorwirth}, {Parise},
  {Schilke}, {Comito}, {Wyrowski}, {G{\"u}sten}, {Bergman}, {Menten}, \&
  {Nyman}}]{leurini2006}
{Leurini}, S., {Rolffs}, R., {Thorwirth}, S., {et~al.} 2006, \aap, 454, L47

\bibitem[{{Lis} \& {Schilke}(2003)}]{lis2003}
{Lis}, D.~C. \& {Schilke}, P. 2003, \apjl, 597, L145

\bibitem[{{Menten} {et~al.}(2007){Menten}, {Reid}, {Forbrich}, \&
  {Brunthaler}}]{menten2007}
{Menten}, K.~M., {Reid}, M.~J., {Forbrich}, J., \& {Brunthaler}, A. 2007, \aap,
  474, 515

\bibitem[{{Monje} {et~al.}(2011{\natexlab{a}}){Monje}, {Emprechtinger},
  {Phillips}, {Lis}, {Goldsmith}, {Bergin}, {Bell}, {Neufeld}, \&
  {Sonnentrucker}}]{monje2011}
{Monje}, R.~R., {Emprechtinger}, M., {Phillips}, T.~G., {et~al.}
  2011{\natexlab{a}}, \apjl, 734, L23

\bibitem[{{Monje} {et~al.}(2011{\natexlab{b}}){Monje}, {Phillips}, {Peng},
  {Lis}, {Neufeld}, \& {Emprechtinger}}]{monje2011clover}
{Monje}, R.~R., {Phillips}, T.~G., {Peng}, R., {et~al.} 2011{\natexlab{b}},
  \apjl, 742, L21

\bibitem[{{Neufeld} {et~al.}(2006){Neufeld}, {Schilke}, {Menten}, {Wolfire},
  {Black}, {Schuller}, {M{\"u}ller}, {Thorwirth}, {G{\"u}sten}, \&
  {Philipp}}]{neufeld2006}
{Neufeld}, D.~A., {Schilke}, P., {Menten}, K.~M., {et~al.} 2006, \aap, 454, L37

\bibitem[{{Neufeld} {et~al.}(2010){Neufeld}, {Sonnentrucker}, {Phillips},
  {Lis}, {de Luca}, {Goicoechea}, {Black}, {Gerin}, {Bell}, {Boulanger},
  {Cernicharo}, {Coutens}, {Dartois}, {Kazmierczak}, {Encrenaz}, {Falgarone},
  {Geballe}, {Giesen}, {Godard}, {Goldsmith}, {Gry}, {Gupta}, {Hennebelle},
  {Herbst}, {Hily-Blant}, {Joblin}, {Ko{\l}os}, {Kre{\l}owski},
  {Mart{\'{\i}}n-Pintado}, {Menten}, {Monje}, {Mookerjea}, {Pearson},
  {Perault}, {Persson}, {Plume}, {Salez}, {Schlemmer}, {Schmidt}, {Stutzki},
  {Teyssier}, {Vastel}, {Yu}, {Cais}, {Caux}, {Liseau}, {Morris}, \&
  {Planesas}}]{neufeld2010}
{Neufeld}, D.~A., {Sonnentrucker}, P., {Phillips}, T.~G., {et~al.} 2010, \aap,
  518, L108

\bibitem[{{Neufeld} \& {Wolfire}(2009)}]{neufeld2009}
{Neufeld}, D.~A. \& {Wolfire}, M.~G. 2009, \apj, 706, 1594

\bibitem[{{Phillips} {et~al.}(2010){Phillips}, {Bergin}, {Lis}, {Neufeld},
  {Bell}, {Wang}, {Crockett}, {Emprechtinger}, {Blake}, {Caux}, {Ceccarelli},
  {Cernicharo}, {Comito}, {Daniel}, {Dubernet}, {Encrenaz}, {Gerin}, {Giesen},
  {Goicoechea}, {Goldsmith}, {Herbst}, {Joblin}, {Johnstone}, {Langer},
  {Latter}, {Lord}, {Maret}, {Martin}, {Melnick}, {Menten}, {Morris},
  {M{\"u}ller}, {Murphy}, {Ossenkopf}, {Pearson}, {P{\'e}rault}, {Plume},
  {Qin}, {Schilke}, {Schlemmer}, {Stutzki}, {Trappe}, {van der Tak}, {Vastel},
  {Yorke}, {Yu}, {Zmuidzinas}, {Boogert}, {G{\"u}sten}, {Hartogh}, {Honingh},
  {Karpov}, {Kooi}, {Krieg}, \& {Schieder}}]{phillips2010}
{Phillips}, T.~G., {Bergin}, E.~A., {Lis}, D.~C., {et~al.} 2010, \aap, 518,
  L109

\bibitem[{{Pickett} {et~al.}(1998){Pickett}, {Poynter}, {Cohen}, {Delitsky},
  {Pearson}, \& {M{\"u}ller}}]{pickett1998}
{Pickett}, H.~M., {Poynter}, R.~L., {Cohen}, E.~A., {et~al.} 1998, \jqsrt, 60,
  883

\bibitem[{{Pilbratt} {et~al.}(2010){Pilbratt}, {Riedinger}, {Passvogel},
  {Crone}, {Doyle}, {Gageur}, {Heras}, {Jewell}, {Metcalfe}, {Ott}, \&
  {Schmidt}}]{pilbratt2010}
{Pilbratt}, G.~L., {Riedinger}, J.~R., {Passvogel}, T., {et~al.} 2010, \aap,
  518, L1

\bibitem[{{Rangwala} {et~al.}(2011){Rangwala}, {Maloney}, {Glenn}, {Wilson},
  {Rykala}, {Isaak}, {Baes}, {Bendo}, {Boselli}, {Bradford}, {Clements},
  {Cooray}, {Fulton}, {Imhof}, {Kamenetzky}, {Madden}, {Mentuch}, {Sacchi},
  {Sauvage}, {Schirm}, {Smith}, {Spinoglio}, \& {Wolfire}}]{rangwala2011}
{Rangwala}, N., {Maloney}, P.~R., {Glenn}, J., {et~al.} 2011, \apj, 743, 94

\bibitem[{{Reese} {et~al.}(2005){Reese}, {Stoecklin}, {Voronin}, \&
  {Rayez}}]{reese2005}
{Reese}, C., {Stoecklin}, T., {Voronin}, A., \& {Rayez}, J.~C. 2005, \aap, 430,
  1139

\bibitem[{Roelfsema {et~al.}(2011)Roelfsema, Helmich, Teyssier, 3, 4, 5, 6, 7,
  \& 8}]{roelfsema2011}
Roelfsema, P., Helmich, F., Teyssier, D., {et~al.} 2011, \aap, in press

\bibitem[{{Sch{\"o}ier} {et~al.}(2005){Sch{\"o}ier}, {van der Tak}, {van
  Dishoeck}, \& {Black}}]{schoeier2005}
{Sch{\"o}ier}, F.~L., {van der Tak}, F.~F.~S., {van Dishoeck}, E.~F., \&
  {Black}, J.~H. 2005, \aap, 432, 369

\bibitem[{{Simon} {et~al.}(1997){Simon}, {Stutzki}, {Sternberg}, \&
  {Winnewisser}}]{simon1997}
{Simon}, R., {Stutzki}, J., {Sternberg}, A., \& {Winnewisser}, G. 1997, \aap,
  327, L9

\bibitem[{{Sofia} {et~al.}(2011){Sofia}, {Parvathi}, {Babu}, \&
  {Murthy}}]{sofia2011}
{Sofia}, U.~J., {Parvathi}, V.~S., {Babu}, B.~R.~S., \& {Murthy}, J. 2011, \aj,
  141, 22

\bibitem[{{Sonnentrucker} {et~al.}(2010){Sonnentrucker}, {Neufeld}, {Phillips},
  {Gerin}, {Lis}, {de Luca}, {Goicoechea}, {Black}, {Bell}, {Boulanger},
  {Cernicharo}, {Coutens}, {Dartois}, {Ka{\'z}mierczak}, {Encrenaz},
  {Falgarone}, {Geballe}, {Giesen}, {Godard}, {Goldsmith}, {Gry}, {Gupta},
  {Hennebelle}, {Herbst}, {Hily-Blant}, {Joblin}, {Ko{\l}os}, {Kre{\l}owski},
  {Mart{\'{\i}}n-Pintado}, {Menten}, {Monje}, {Mookerjea}, {Pearson},
  {Perault}, {Persson}, {Plume}, {Salez}, {Schlemmer}, {Schmidt}, {Stutzki},
  {Teyssier}, {Vastel}, {Yu}, {Caux}, {G{\"u}sten}, {Hatch}, {Klein}, {Mehdi},
  {Morris}, \& {Ward}}]{sonnentrucker2010}
{Sonnentrucker}, P., {Neufeld}, D.~A., {Phillips}, T.~G., {et~al.} 2010, \aap,
  521, L12

\bibitem[{{St{\"o}rzer} {et~al.}(1995){St{\"o}rzer}, {Stutzki}, \&
  {Sternberg}}]{stoerzer1995}
{St{\"o}rzer}, H., {Stutzki}, J., \& {Sternberg}, A. 1995, \aap, 296, L9

\bibitem[{{Stutzki} {et~al.}(1988){Stutzki}, {Stacey}, {Genzel}, {Harris},
  {Jaffe}, \& {Lugten}}]{stutzki1988}
{Stutzki}, J., {Stacey}, G.~J., {Genzel}, R., {et~al.} 1988, \apj, 332, 379

\bibitem[{{Tauber} {et~al.}(1994){Tauber}, {Tielens}, {Meixner}, \&
  {Foldsmith}}]{tauber1994}
{Tauber}, J.~A., {Tielens}, A.~G.~G.~M., {Meixner}, M., \& {Foldsmith}, P.~F.
  1994, \apj, 422, 136

\bibitem[{{Th{\"u}mmel} {et~al.}(1992){Th{\"u}mmel}, {Nesbet}, \&
  {Peyerimhoff}}]{thuemmel1992}
{Th{\"u}mmel}, H.~T., {Nesbet}, R.~K., \& {Peyerimhoff}, S.~D. 1992, Journal of
  Physics B Atomic Molecular Physics, 25, 4553

\bibitem[{{Van der Tak}(2011)}]{vdtak2011}
{Van der Tak}, F.~F.~S. 2011, in IAU Symposium, Vol. 280, arXiv:1107.3368

\bibitem[{{Van der Tak} {et~al.}(2007){Van der Tak}, {Black}, {Sch{\"o}ier},
  {Jansen}, \& {van Dishoeck}}]{vdtak2007}
{Van der Tak}, F.~F.~S., {Black}, J.~H., {Sch{\"o}ier}, F.~L., {Jansen}, D.~J.,
  \& {van Dishoeck}, E.~F. 2007, \aap, 468, 627

\bibitem[{{Van der Werf} {et~al.}(2010){Van der Werf}, {Isaak}, {Meijerink},
  {Spaans}, {Rykala}, {Fulton}, {Loenen}, {Walter}, {Wei{\ss}}, {Armus},
  {Fischer}, {Israel}, {Harris}, {Veilleux}, {Henkel}, {Savini}, {Lord},
  {Smith}, {Gonz{\'a}lez-Alfonso}, {Naylor}, {Aalto}, {Charmandaris}, {Dasyra},
  {Evans}, {Gao}, {Greve}, {G{\"u}sten}, {Kramer}, {Mart{\'{\i}}n-Pintado},
  {Mazzarella}, {Papadopoulos}, {Sanders}, {Spinoglio}, {Stacey}, {Vlahakis},
  {Wiedner}, \& {Xilouris}}]{vdwerf2010}
{Van der Werf}, P.~P., {Isaak}, K.~G., {Meijerink}, R., {et~al.} 2010, \aap,
  518, L42

\bibitem[{{Van der Wiel} {et~al.}(2009){Van der Wiel}, {van der Tak},
  {Ossenkopf}, {Spaans}, {Roberts}, {Fuller}, \& {Plume}}]{vdwiel2009}
{Van der Wiel}, M.~H.~D., {van der Tak}, F.~F.~S., {Ossenkopf}, V., {et~al.}
  2009, \aap, 498, 161

\bibitem[{{Wakelam} \& {Herbst}(2008)}]{wakelam2008}
{Wakelam}, V. \& {Herbst}, E. 2008, \apj, 680, 371

\bibitem[{{Walmsley} {et~al.}(2000){Walmsley}, {Natta}, {Oliva}, \&
  {Testi}}]{walmsley2000}
{Walmsley}, C.~M., {Natta}, A., {Oliva}, E., \& {Testi}, L. 2000, \aap, 364,
  301

\bibitem[{{Wang} {et~al.}(1993){Wang}, {Jaffe}, {Evans}, {Hayashi},
  {Tatematsu}, \& {Zhou}}]{wang1993}
{Wang}, Y., {Jaffe}, D.~T., {Evans}, II, N.~J., {et~al.} 1993, \apj, 419, 707

\bibitem[{{Young Owl} {et~al.}(2000){Young Owl}, {Meixner}, {Wolfire},
  {Tielens}, \& {Tauber}}]{youngowl2000}
{Young Owl}, R.~C., {Meixner}, M.~M., {Wolfire}, M., {Tielens}, A.~G.~G.~M., \&
  {Tauber}, J. 2000, \apj, 540, 886

\bibitem[{{Zhang} {et~al.}(2009){Zhang}, {Faure}, \& {Tennyson}}]{zhang2009}
{Zhang}, R., {Faure}, A., \& {Tennyson}, J. 2009, \physscr, 80, 015301

\end{thebibliography}

\end{document}